\documentclass[11pt,twoside]{article}
\usepackage{asp2004}
\usepackage{psfig}
\usepackage{epsf}
\usepackage{graphics}
\usepackage{lscape}
\def\edcomment#1{\iffalse\marginpar{\raggedright\sl#1\/}\else\relax\fi}
\reversemarginpar

\newcommand{\mdot}{M$_{\odot}$}

\begin{document}
\title{Survival Rates and Consequences}
\author{B. C. Whitmore}
\affil{Space Telescope Science Institute, 3700 San Martin Dr., Baltimore, MD, 21218, USA}

\begin{abstract}
To first order, the initial cluster luminosity function appears to be
universal. This means that the brightest young cluster in a galaxy can
be predicted from the total number of young clusters based purely on
statistics. This suggests that the physical processes responsible for the
formation of clusters are similar in a wide variety of galaxies,
from mergers to quiescent spirals. One possibility is that conditions for making young massive
clusters are globally present in mergers while only locally present in
spirals (i.e., in the spiral arms). 
However, understanding the
destruction of clusters and the accompanying survival rates is
more important for understanding cluster demographics than understanding
their formation. This is because only about 1 in
1,000 clusters with mass greater than 10$^4$ \mdot\/
will survive to become an old globular cluster.  In this
paper we briefly review this basic framework
and then develop a toy model
that allows us to begin to address several fundamental
questions.  In particular, we demonstrate that young clusters in the
Antennae Galaxies have a high
``infant mortality'' rate, with roughly 90 \% of the clusters being
destroyed each decade of log(time). We also advocate the use of an
objective classification system for clusters, with the three
parameters being mass, age, and size.

\end{abstract}
\thispagestyle{plain}

\section{Introduction}
Historically, star clusters have been divided into three types 
in the Milky Way; globular clusters, open clusters, and associations.
This has conditioned us to think in terms of three distinct modes of cluster
formation. However, when we look at external galaxies we see
a power law continuum of young cluster masses. How can these two
outlooks be reconciled?

There is growing evidence  that all star clusters may form from a
universal initial cluster mass function, which is then modified by
a variety of destruction mechanisms. These destruction mechanisms include
infant mortality (e.g., unbound clusters that freely expand in the
first $\approx$ 10 Myr), environmental effects (e.g., disk and bulge
shocking, dynamical friction), internal effects (e.g., evaporation from
2-body relaxation )
and stellar evolution (e.g., mass loss). By convolving these
formation and destruction rates with
the star formation history of a galaxy, and after taking
into effect various observational artifacts and selection effects,
we should be able to predict the current distribution of young and old star
clusters we see in a particular galaxy.

In this contribution we first outline some of the results which
have lead to the development of
this basic framework, and then develop a ``toy'' model
designed to help answer three fundamental questions.

\subsection{Is the Initial Cluster Mass Function Continuous or Modal ?} 

Essentially all studies of the luminosity functions of young clusters
have found them to be power laws,
 $\phi(L)dL \propto L^{\alpha} dL$ with a value of $\alpha$ $\approx$ --2 
(e.g., Whitmore et al. 1999, see Whitmore 2003 for a review [originally
appearing in 2000 as astro-ph/0012546]). Mass
functions have been determined for only a small subset of these
galaxies, but a similar power law relationship is generally found,
again with an index --2 (e.g., Zhang and Fall 1999). On the other
hand, the luminosity and mass functions of old globular clusters
are peaked, with a mean $V$ magnitude
M$_v$ $\approx$  -7.4, a mean mass $\approx$ 2 $\times$ 10$^5$ \mdot, 
and a width
$\sigma \approx 1.4$ mag (e.g., Whitmore 2003). Several
theoretical studies (e.g., Vesperini 1998, Fall and Zhang 2001) have
suggested that a natural way to explain this apparent discrepancy
is the destruction of the fainter, less massive clusters, due to
effects such as 2-body relaxation, tidal shocks, and stellar mass loss.

Other theoretical studies have also highlighted
the important role that cluster destruction may play in determining
the demographics of clusters (e.g., Fall \& Rees 1977, Gnedin \& Ostriker 1997).
The current paper adopts a similar framework, but broadens it to
include all star clusters; young and old, near and far. The basic
question is: {\it ``Can all observations of star cluster demographics
be explained by a universal
initial cluster mass function followed by the destruction
of a subset of the clusters which carves away regions of 
parameter space ?''}

\subsection{Is the Initial Cluster Mass Function Universal ?}

On the surface, the demographics of star clusters appear
to differ dramatically in different galaxies.  For example, merging galaxies
have large numbers of very bright, very massive young clusters, often
called super star clusters. On the other hand, in relatively quiescent
galaxies, such as the Milky Way, the young clusters appear to be primarily 
faint and 
low mass. We might conclude that
certain types of clusters can only be formed in certain types of 
galaxies. This will be referred to as ``special creation'' in this paper.


An alternative to ``special creation'' is suggested by the work of 
Elmegreen and Efremov (1997),
Whitmore (2003), and Larsen (2002), who all suggest that the initial
cluster mass function may be universal.
Whitmore (2003) examined the luminosity functions of eight galaxies as
part of his review of the formation of star clusters. Part of the
motivation was to look for a truncation at the bright end of the luminosity
function  which might
indicate that only the major mergers can
produce the very massive clusters (i.e., supporting ``special creation''). 
However, he found that the cluster
luminosity functions  for mergers, starbursts,
and barred galaxies were all power laws.
The main difference is the
normalization of the power law, with young active mergers having
thousands of young clusters, while other systems  have
hundreds of young clusters.  The large difference in numbers might 
indicate that
conditions for making clusters are {\it globally} present in mergers
but only {\it locally}
present (e.g., in spiral arms) in more quiescent galaxies.


Since the number of galaxies with sufficient data to form a
meaningful luminosity function was limited, Whitmore (2003) went on
to use a poor-man's
version of the diagram, shown in Figure 1, to enlarge the sample.  This
includes the sample of spiral galaxies from Larsen \& Richtler (2000)
and plots only the luminosity of the brightest cluster vs. the log of the
number of
clusters in a galaxy that are brighter than M$_V$ = --8. Somewhat
surprisingly, this shows no

\clearpage 

\begin{figure}[hb]	
\plotfiddle{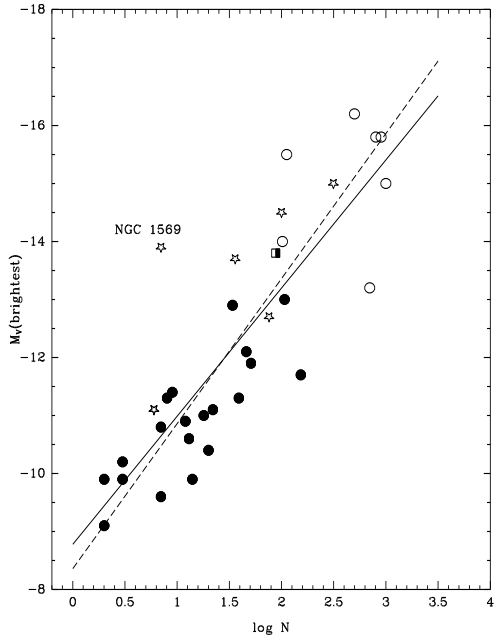}{5.5 cm}{0}{35}{35}{-100}{-80}
\vspace{2 cm}
\caption{Magnitude of the brightest cluster vs. log of the number of
clusters in 34 galaxies. Filled circles are spiral galaxies from
Larsen \& Richtler (2000), open circles are mergers, stars are
starbursting galaxies, and the half-filled square is a barred
galaxy. The solid line is a best fit (excluding NGC 1569) while the
dashed line is the prediction from a universal power law luminosity
function with index $\alpha$ = --2. See Whitmore (2003) for details.}
\end{figure}

\noindent evidence of a truncation in the ability of
quiescent spiral galaxies to form bright star clusters. The observed
slope is consistent with nearly all of the galaxies having the same
universal luminosity function. This suggests that active mergers have the brightest
clusters only because they have the most clusters.  {\it It appears to
be a matter of simple statistics rather than a difference in
physical formation mechanisms.}

Larsen (2002) has further developed this basic idea by performing
Monte-Carlo simulations that show that the spread in the
scatter of the relation can also be explained by statistics.
Several other recent studies have also begun to consider
this ``size-of-sample'' effect   (e.g., Hunter et al., 2003, and  
Billet, Hunter, \& Elmegreen 2002).

\subsection{What Fraction of Stars are Formed in Clusters and What 
Fraction are Formed in the Field ?}

Understanding the destruction of star clusters is more important for 
understanding their demographics than understanding how they form.
For every cluster that survives to an age of 10 Gyr, roughly a
thousand were created 
and have been destroyed, their stars being dispersed into the field.
 As a result of this, 
it now appears that the majority of stars are created in clusters
rather than in the field.

Whitmore (2003) points out that if we assume that the Antennae has been
making clusters at the same rate for the past 200 Myr (a rather
uncertain assumption), it is possible to use Figure 2 from Zhang and Fall 
(1999, an updated figure is also shown as Figure 2c below),  to show
that for every 20 clusters in the age range 0 - 10 Myr, only one will survive to
an age of $\approx$ 100 Myr (i.e., there are roughly the same number
of clusters in the 0 - 10 Myr age bin as in the 20 - 200 Myr age bin).
While this crude calculation is probably not justifiable for a single
galaxy, larger samples are now becoming available (e.g., M83 - Harris et al. 2001, LMC - Hunter et al. 2003, NGC 6745 - de Grijs et al. 2003, 
    M51 and M101 - Chandar et al. 2004)
which show similar demographics in nearly all star forming
galaxies that have been observed.

\section{A Framework for Understanding Star Cluster Demographics} 

In this section we outline our framework for understanding the
demographics of star clusters. The key ingredients are:

\begin{itemize}

\item A universal initial mass function (i.e., power law with index --2).

\item Variable star formation histories (e.g., continuous or bursts; Bruzual-Charlot
2003 models are used for the SED models).

\item Variable cluster disruption mechanisms (3 models have currently been
enabled, constant mass loss, an empirical formula from Boutloukous \& Lamers   (2003), and ``infant mortality'' (i.e., removal of 90 \% of the clusters
every decade of log(time)
for the first 100 Myr).

\item Convolution with observational artifacts and selection effects
(e.g., magnitude thresholds, reddening, extinction,
artifacts from age-dating algorithms).

\end{itemize}

The goal is to simulate and explain a wide range of observations
for a wide variety of different galaxies (e.g., luminosity functions, age histograms, size histograms, color-mag, 
color-color diagrams, ...).

It is easy to  think of several apparent exceptions to
the framework outlined above. For example, 
where are the intermediate-age globular
clusters in the Milky Way? Doesn't the existence of a very bright
cluster in the dwarf galaxy NGC 1569 (Figure 1) argue for ``special 
formation'', at least for this single case?

The approach we plan to take as we develop this model in future papers
is to adopt this simple framework to see
how far we can go with it. Can it explain the basic trends
but not a few exceptions? Is it actually possible to explain even these apparent
exceptions given a little more thought? For example, perhaps NGC 1569
represents the 2 sigma outlier that is sure to be observed
while the galaxies without bright clusters, that would fall below the
line, have not yet been observed.

Another possible exception is the absence of intermediate-age
globular clusters in the disk of the Milky Way. Perhaps this indicates
that there is an upper limit to the mass of GMCs in the Milky Way. Or
perhaps there is a destruction mechanism that is unique to the disk
environment (e.g., tidal shocking from GMCs). On the other hand,
perhaps intermediate-age globular clusters do exist in the Milky Way
(as they do in other nearby galaxies such as M31- Barmby 2002 and M33 -
Chandar et al. 1999) and will be found in several IR searches of the Milky Way disk that are
now in progress.

\section{A Toy Model - Early Insights}

We have developed a simple toy model to determine whether the 
framework outlined above can explain various observations.
Even at the early stages of development represented by the present
contribution a number of insights into the formation and destruction of
star clusters are apparent, as described below.

\subsection{Insight 1 - Expectations from Continuous Cluster Formation}

Figures 2a (upper left)  shows the mass vs. age diagram for a model with
continuous cluster formation and an initial mass distribution which is
a power law with index $\alpha$ = --2. 
Figure 2b (upper right) shows the same simulation with a magnitude
cutoff imposed.
One nice attribute of this model is that
the number of clusters in a decade of mass increase by a factor of 10,
hence there is equal mass in each decade (Lada \& Lada 2003).  Similarly, there is a
factor of 10 more clusters in each decade of time, by definition.

Figures 2a and 2b provides a good opportunity for demonstrating the
size-of-samples effect. Taken at face value, it
appears that massive clusters were only produced in the past, since the most
massive young clusters are roughly 10$^4$ \mdot. However, the clusters
{\it were all taken from the same power law distribution function};
the only reason for the lack of massive young clusters is the factor
of 1000 fewer clusters in the 5-10 Myr bin as compared to the 5-10 Gyr
bin.  Hence, statistics rather than physics are responsible for the
lack of massive young clusters. There are simply not enough clusters
in the sample to produce a 2 sigma statistical deviation required to
produce a 10$^5$ \mdot\/ cluster, let alone a 3 sigma deviation required
for a 10$^6$ \mdot\/ cluster.

\subsection{Insight  2 - The Need for Infant Mortality}

The continuous star formation model shown in Figures 2a and 2b  look
much different than the age distribution for clusters in the Antennae
galaxies, shown in Figure 2c (bottom left).  Instead of the factor of
10 increase in each decade of time predicted by continuous cluster
formation,  we find roughly
equal numbers of clusters in each age bin (staying above the completeness
threshold  at 10$^{4.5}$ \mdot\/
with log(time) $<$ 9).  This immediately tells us that if the Antennae
has had roughly continuous cluster formation, clusters must be
destroyed at a rate of time$^{-1}$ to offset the increase in bin size
which is proportional to time.  Similar arguments have been made by
others (e.g., Whitmore 2003).  A more detailed discussion is included
by Fall (2004) and Fall, Chandar, \& Whitmore (2004).  

Perhaps the
Antennae had a recent burst of cluster formation. This is likely to be
true, since regions like the overlap region have large numbers of very
young clusters. However, if we look at each of the four WFPC2 chips
independently (Figure 3), we find that all of them look quite similar,
with the largest numbers of clusters always having ages $<$ 10$^7$
years !
Since it is not possible for the galaxy to synchronize its burst of
star formation over its entire disk (i.e., assuming a ``sound speed'' of
$\approx$ 30 km s$^{-1}$ [Whitmore et al. 1999] requires approximately 300 Myr for one side of the
galaxy to communicate with the other side 10 kpc away), we can only
conclude that the vast majority of the clusters are being destroyed
almost as fast as they form, making it appear that the largest number
of clusters are always the youngest.

\clearpage

\begin{figure}[hb]	
\plotfiddle{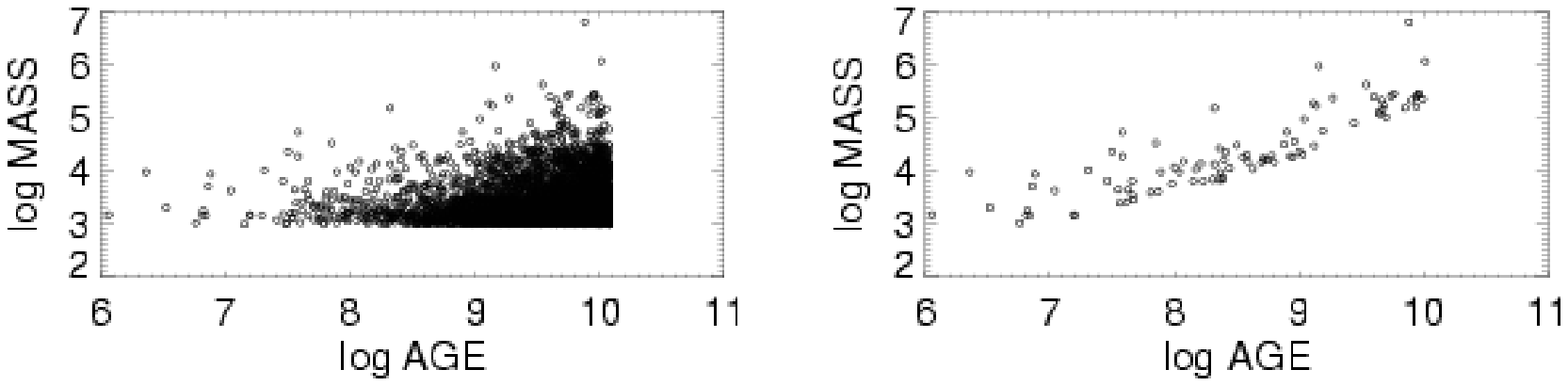}{1.5 cm}{0}{90}{90}{-210}{-350}
\end{figure}

\begin{figure}[hb]	
\plotfiddle{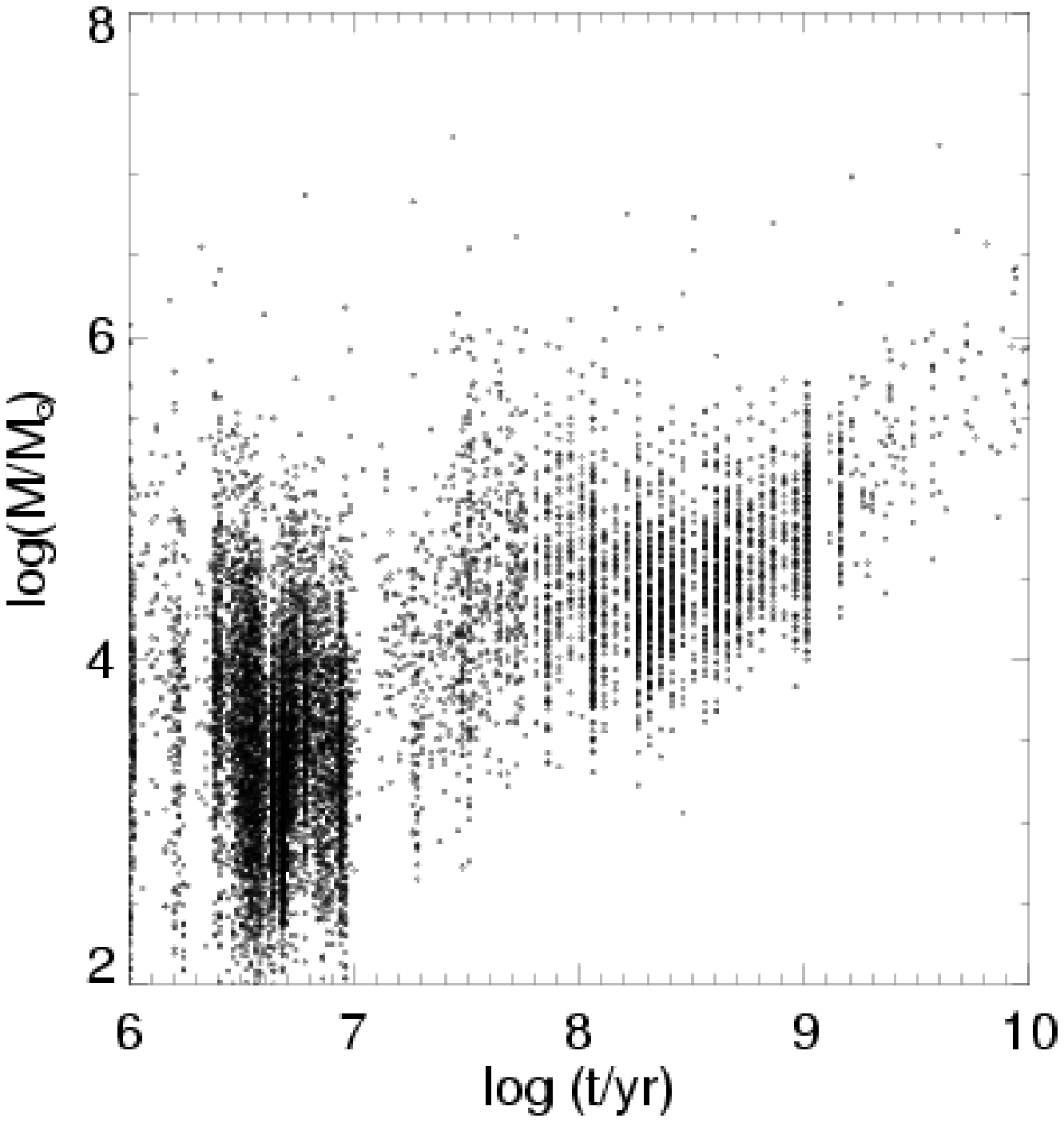}{6.0 cm}{0}{50}{50}{-250}{-160}  
\end{figure}

\begin{figure}[hb]	
\plotfiddle{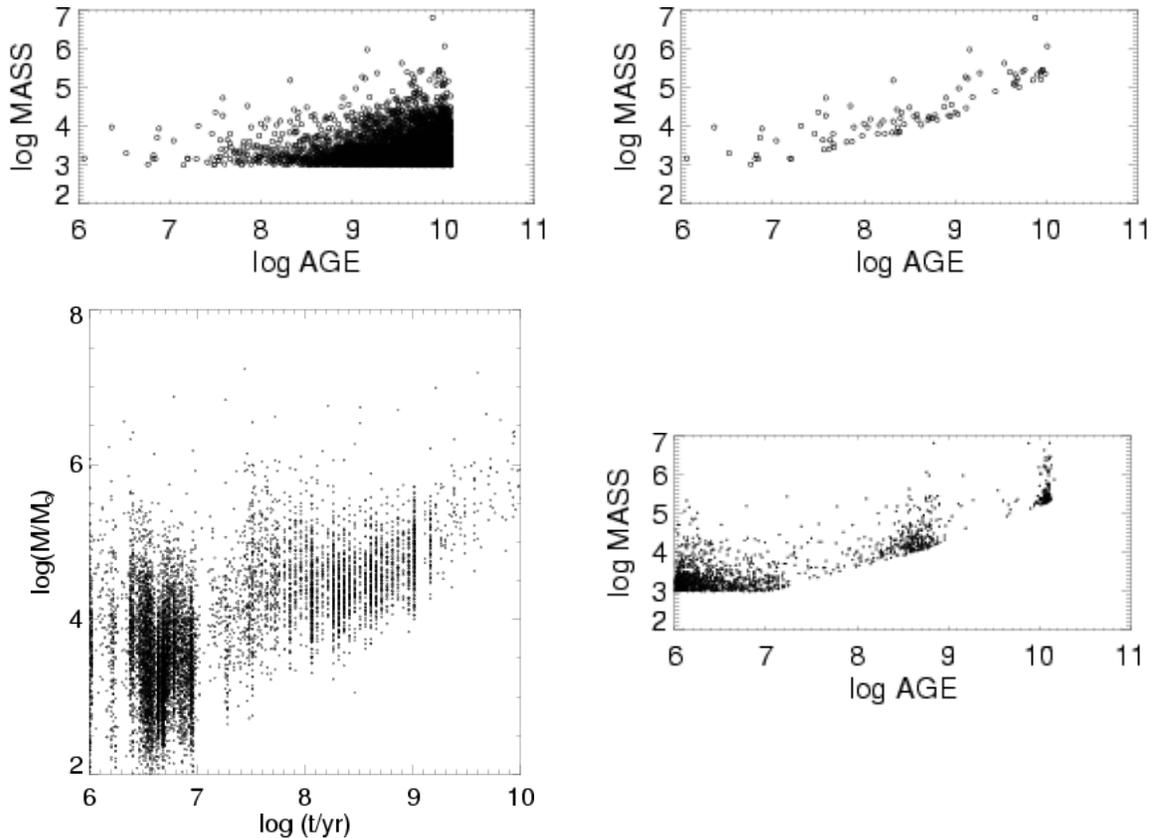}{2cm}{0}{90}{90}{20}{80}
\vspace {-1 cm}
\caption{2a (upper left) - Continuous formation model; 2b (upper right)
continuous formation model convolved with a magnitude threshold; 2c (lower
left) - observed data from the Antennae (Fall, Chandar, Whitmore 2004);
2d (lower right) early version of toy model for the 
Antennae with continuous formation for the last 10$^{10}$ years, 
an additional
continuous formation ``burst'' for the last 100 Myr,  
Gaussian bursts at 500 Myr and 12 Gyr, and infant mortality plus constant mass loss destruction laws. Artifacts include the 10$^3$ \mdot\/ cutoff artifact for the simulations, and the apparent gap around 10 Myr in the Antennae
observations (due to age-dating difficulties).}

\end{figure}

If one looks closely at Figure 3  there actually are small
differences on different chips (e.g., the ``overlap''
region portion of WF3 has the largest number of
very young clusters), but these are secondary effects;
the primary correlation is the rapid decline in the number of clusters
as a function of age.

\subsection{Insight 3 - Early Results from the Inclusion of Destruction Laws}

Three heuristic destruction laws have been incorporated into the toy
model at present: 1) constant mass loss (e.g., typical values of
10$^{-5}$ \mdot/yr result in reasonable looking models), 2) the Lamers model 
(e.g., Boutloukous \& Lamers, 2003), 
and 3) infant mortality (parameterized in this simple model
as random removal of 90 \% of the
clusters every decade of log(time) for the first 100 Myrs).

\clearpage
\begin{figure}[hb]	
\plotfiddle{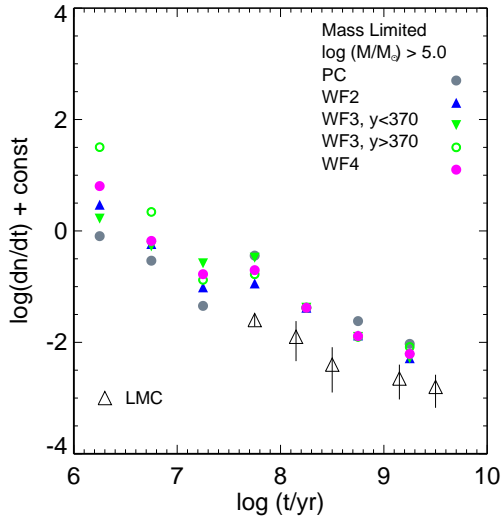}{5.5cm}{0}{40}{40}{-120}{-50}
\vspace{0.5 cm}
\caption{The log(dn/dt) vs. log(time) diagram for the Antennae with results for the four WFPC2 chips shown separately (normalized at log(age) = 8.2). The WF3 chip is broken into two regions; one
being the very active ``overlap'' region (y $>$ 370), the other
being a region outside the overlap region in an area 
with a high proportion of older clusters (y $<$ 370). The open
triangles show the trends in the LMC (Elson \& Fall 1985).}
\end{figure}


We note that the Lamers model with
values of T4=100 Myr (appropriate for their results on M33
and M51) predicts that even a cluster with 10$^{6}$ \mdot\/ is
destroyed in $\approx$ 1 Gyr, implying that no young clusters
evolve to become old globular clusters !  We also note that the Lamers
formula predicts essentially no destruction occurs until about 100
Myr, implying that all young clusters are
bound and that the infant
mortality rate is zero. These are likely to be artifacts
of an empirically determined model fit over a limited age range.
Care must be used to only use this model in the appropriate
age range.

\section{The Antennae as a Scaled-Up Version of the Milky Way}

Lada \& Lada (2003) make
several of the same points about the need for infant mortality for the
Milky Way that are made for the Antennae in this paper, even 
to the level of giving the same numbers (i.e., 
90 \% destruction after 10 Myr)! 
In fact, it is possible to view the Antennae as a scaled-up version of
the Milky Way. According to Lada \& Lada (2003), the Milky Way has roughly
100 known embedded young clusters within a radius of $\approx$ 2 kpc. The
most massive clusters in this sample are slightly larger than 
10$^3$ \mdot.  If we
were able to see the entire disk of the Milky Way the sample would be
roughly 100 times larger. The star formation rate per unit area in the Antennae
is also roughly a factor of 10 higher than the Milky Way (Zhang, Fall, \& Whitmore 2003)
hence the total enhancement for the number in the sample
might be roughly a factor
of 1000. Assuming a universal power law
with index --2 for the initial mass function, the most massive clusters 
with comparable ages to the Lada \& Lada sample (i.e., $\approx$ 10 Myr) 
would be predicted to be $\approx$ 10$^6$ \mdot, just
as they are in the Antennae.
 
\section{An Objective Classification System for Star Clusters} 

Several authors (e.g., Whitmore 2003, Terlevich 2004)
have pointed out the large number of different
names used to describe basically similar objects (i.e., super star clusters,
young massive clusters, populous clusters, young globular clusters, ...).
Conversely, other authors (e.g., Hodge 1988, see Figure 2 for a dramatic
example) have pointed out how dissimilar objects that we call globular
can be.

It may
be beneficial to consider an objective
classification system that would help unify the 
situation.
Such a system would also provide a
more meaningful method of comparing clusters in different
galaxies, near and far.

Hodge (1988) suggested an objective 3-dimensional classification
system for star clusters consisting of mass, age, and
metallicity. While the first two parameters are clearly fundamental, 
metallicity
appears to be less important in a dynamical sense and is largely correlated with 
age in any case. In the context of our statement that understanding the
destruction of clusters will be critical to understanding their
demographics, a better choice for the third parameter might be the
density or size of a cluster, which largely controls how susceptible a
cluster is to destruction. As pointed out by Terlevich (private
communication), size would be
the better choice since density is a derived quantity.


\end{document}